\documentclass[conference]{IEEEtran}
\usepackage{cite}
\usepackage{amsmath,amssymb,amsfonts}
\usepackage{algorithmic}
\usepackage{graphicx}
\usepackage{textcomp}
\usepackage{enumerate}
\usepackage{epstopdf}
\usepackage[table, svgnames, dvipsnames]{xcolor}
\usepackage{xcolor}
\def\BibTeX{{\rm B\kern-.05em{\sc i\kern-.025em b}\kern-.08em
    T\kern-.1667em\lower.7ex\hbox{E}\kern-.125emX}}

\begin{document} 

\title{A Block Diagonal Markov Model for Indoor Software-Defined Power Line Communication\\}

\author{\IEEEauthorblockN{Ayokunle Damilola FAMILUA}
\IEEEauthorblockA{\textit{Center for Telecommunications,} \\
\textit{Department of Electrical and Electronics Engineering Science}\\
\textit{University of Johannesburg}\\
Johannesburg, South Africa \\
familuaad@uj.ac.za}}

% \IEEEoverridecommandlockouts
% \IEEEpubid{\makebox[\columnwidth]{978-1-7281-1010-3/19/\$31.00~\copyright2019 IEEE \hfill} \hspace{\columnsep}\makebox[\columnwidth]{ }}

\maketitle

\thispagestyle{plain}
\pagestyle{plain}
% \IEEEpubidadjcol

\begin{abstract}
A Semi-Hidden Markov Model (SHMM) for bursty error channels is defined by a state transition probability matrix $A$, a prior probability vector $\Pi$, and the state dependent output symbol error probability matrix $B$. Several processes are utilized for estimating $A$, $\Pi$ and $B$ from a given empirically obtained or simulated error sequence. However, despite placing some restrictions on the underlying Markov model structure, we still have a computationally intensive estimation procedure, especially given a large error sequence containing long burst of identical symbols. Thus, in this paper, we utilize under some moderate assumptions, a Markov model with random state transition matrix $A$ equivalent to a unique \emph{Block Diagonal Markov model} with state transition matrix $\Lambda$ to model an indoor software-defined power line communication system. A computationally efficient modified \emph{Baum-Welch algorithm} for estimation of $\Lambda$ given an experimentally obtained error sequence from the indoor PLC channel is utilized. Resulting Equivalent \emph{Block Diagonal Markov models} assist designers to accelerate and facilitate the procedure of novel PLC systems design and evaluation.
\end{abstract}

\begin{IEEEkeywords}
Baum-Welch Algorithm, Block Diagonal Markov Model, PLC, Smart Grid, Software defined NB-PLC
\end{IEEEkeywords}

\section{Introduction}
Power line communication (PLC) technology is a wireline communication technology that enables the transmission and reception of data by floating a modulated carrier signal over the existing ubiquitous power line infrastructures. This technology offers a low cost communication technology that requires no new wire installation compared to other wireline communication technologies. Its application across its use in intelligent home system design \cite{b1}, smart home \cite{b2}, grid automation and smart grid applications \cite{b3}, \cite{b4}, \cite{b5}, advanced/smart metering \cite{b6}, street light monitoring and control \cite{b5}, \cite{b7}, power system monitoring and industrial applications (for transmitting control and diagnostic information) \cite{b8} and several other applications.   

The power line infrastructures were not originally conceptualized for data transmission, thus, must overcome several channel impairments in order to achieve an effective and reliable data communication. Factors that inhibit reliable data transmission on the power line includes: impedance mismatch (resulting from ageing cables), utilization of unshielded cables, load dynamics, multipath signal propagation, unpredictability and varying characteristics (time, frequency. location) and noise impairment. The main technical challenges are signal attenuation, signal distortion and noise, with the chief technical challenge being impulse noise which results to burst errors and corruption of transmitted data. Accurate and appropriate channel models that depict these channel impairments assist designers to accelerate and facilitate the procedure of novel PLC systems design and evaluation. Hence, the need for a constant measurement campaign due to the unstable and harsh characteristics of the PLC channel. This measurement campaign are needed from time to time before precise mathematically-based statistical models are derived as: noise parameters are dependent mains voltage, power line topology, time, frequency and location; and PLC exhibits non additive white Gaussian noise (non-AWGN) model contrary to AWGN assumed for several digital communication systems. 

In order to carry out these measurement campaign flexible PLC transceiver systems is the way to go. Greater number of conventional PLC transceiver systems presently available make use of a hardware-based wiring of application specific integrated circuits (ASICs). This approach constrains to a hardware-based architecture modulation schemes and forward error correction techniques essential for a specific standard. Consequently, many of these hardware-based transceivers become obsolete with time as a result of continuous technological advancement and evolution of various communication technologies. Therefore, it is of utmost importance to develop a flexible, reusable, upgradable and reconfigurable PLC transceiver system to cater for measurement campaign in the unstable PLC channel based on a software-defined radio approach. This flexibility will assist designers and communication engineers to ascertain the functionality of adaptive modulation schemes and FEC techniques and observe their performance in exploiting and mitigating noise on the harsh PLC channel. Furthermore, with such approach, drawbacks associated with the conventional hardware-based architecture falls away as one can re-use subparts of an existing transceiver system, and through code reprogramming adapt its operation without the need for hardware architectural changes or replacement.

Thus, in this paper, a single carrier based software-defined narrowband PLC (NB-PLC) transceiver is developed to model an indoor PLC channel using the \emph{Block Diagonal Markov model}, a model based on the Semi-Hidden Markov model (SHMM) approach. A fast and computationally efficient modified \emph{Baum-Welch algorithm} is derived for estimating the parameters of the Block Diagonal Markov model. The implementation is carried out using the MATLAB environment for the software and digital signal processing functions in the transceiver system.

The remaining part of the paper is organized as follows. In Section II, a brief literature of related works is presented. Section III concentrates on the modeling methodology, showing how the Block Diagonal Markov model is constructed in order to model the indoor NB-PLC channel. Notations are presented as well as the construction of the Block Diagonal Markov model underlying mathematics. In Section IV, the modified Baum-Welch algorithm, an iterative algorithm for parameter estimation of the Block Diagonal Markov model parameters is derived. Section V unpacks the development of the single carrier software-defined NB-PLC transceiver system model and its functionality. The model results and analysis are carried out in Section VI. Section VII concludes the paper and gives concluding remarks. 

\section{Related Works}  
Semi-hidden Markov models (SHMMs) a class of hidden Markov model (HMM) based on Markov model underlying principles have been used widely to model burst errors and analyze and simulate  communication systems. The growing interest is focused on the construction of Markov models either from simulation or empirically obtained error sequence and utilizing such models for advance analysis and/or simulation of end-to-end digital communication systems. Application of SHMM or its other variants include its use in burst error modeling in digital channels \cite{b9, b10, b11, b12, b13, b14} speech recognition \cite{b15}, localization in wireless sensor networks \cite{b16}, optimal signal detection in noise \cite{b17}, speech synthesis \cite{b18}, weather prediction, queuing theory, control theory, reliability modeling, and numerous other applications.   

In \cite{b9}, the authors used a first and second-order SHMM based on Fritchman Markov model to model a multi-carrier based indoor NB-PLC channel. Empirical error sequences were obtained and modeled for different multi-carrier modulation schemes at both residential and laboratory site using the conventional Baum-Welch algorithms. Results analysis showed precise channel models obtained which were validated using the log-likelihood ratio plots, the error-free run distribution (EFRD) plots, the mean square error (MSE) and Chi-Square test. The estimated second-order models were ascertained to be more precise models compared to the first-order ones. In \cite{b10}, a first-order semi-hidden Fritchman Markov model (SHFMM) of impulse noise, narrowband noise and background noise was carried out in a typical South African residence and lab based on signal level measurement. A mildly disturbed and heavily disturbed noise scenarios were considered with the standard Baum-Welch algorithm utilized to estimate the most probable SHFMM parameters that statistically depict the empirically obtained error sequence. The authors in \cite{b11} presented a first and second-order Fritchman Markov modeling of a low complexity hybrid PLC and visible light communication (VLC) system in an indoor environment (residential and laboratory) using the conventional Baum-Welch algorithm. A validation of the superiority of the second-order Fritchman model over the first-order one was also validated. Model results obtained can be utilized in facilitating the design and evaluation of robust modulation schemes and FEC codes to mitigate the effect of noise on the combined channel. The conventional recursive Baum-Welch algorithms utilized in parameter estimation for modeling purposes in \cite{b9, b10, b11, b12} are all computationally intensive due to the long strings of identical symbols in the measured error sequence used as training data. 

On the contrary, in \cite{b13}, Turin proposed and developed a modified Baum-Welch algorithm to reduce the computational complexity of the conventional Baum-Welch res-estimation procedure for error sequences containing long stretch of identical observed symbols. Turin’s approach is based on the use of fast algorithms for matrix exponentiation and analytical expressions for matrix sums corresponding to state durations \cite{b14}. For illustration purposes, Turin applied the algorithm in the estimation of the parameters of a binary error source model utilizing computer simulation obtained error sequence. Turin’s method is particularly applicable to general HMM and is not complex for Markov models that doesn't permit transitions among the same observation symbols \cite{b13}. In this paper, we utilize and adapt a faster modified Baum-Welch algorithm for estimating the parameters of the Block Diagonal Markov model for a NB-PLC channel given empirically obtained error sequences that contains a long stretch of identical symbols as proposed by Srinivas in \cite{b14}. In \cite{b14}, Srinivas showed that based on some mild assumptions, a Markov model with random state transition matrix A is equivalent to a unique ``\emph{Block Diagonal Markov Model}'' with state transition matrix $\Lambda$. A fast and computationally efficient modified Baum-Welch algorithm for estimating  given a set of observation is also derived. Since this model is adopted and adapted for modeling in this paper, a more comprehensive derivation of the Block Diagonal Markov model and the corresponding algorithm for parameter estimation is presented in Section IIIA and IIIB respectively. Furthermore, a presentation of related works based on the use of software-defined approach follows. 

In \cite{b19}, the paper presents an implementation of PLC system for vehicular power line communication (VPLC) utilizing the Universal Software Radio Peripheral version 2 (USRP2) with GNU radio platform. The functionality of the VPLC platform was trialed in an indoor 220V mains low voltage environment, as well as in a VPLC 12V direct current (DC) environment. In \cite{b20}, for a smart home environment, an evaluation of software defined radio (SDR) systems was carried out. The benefits and limitations of utilizing SDR was shown and evaluated. The authors in \cite{b21} designed and tested a PLC-based anti-islanding protection system (otherwise referred to as loss-of-mains (LoM)) for a power distribution grid in Finland based on the flexible SDR platform. This system was proposed to combat islanding condition, which is an occurrence in an electricity distribution grid whereby a distributed generator keeps powering parts of the grid, after a loss of connection with the power utility. Laboratory tests were carried, with elaborate analysis of communication-based LoM/anti-islanding protection technologies and solutions presented. In \cite{b22}, Otterbach et al. gave basic PLC details and showed how USRPs can be interfaced with the power supply network for several applications. They further presented the benefits and demo applications (GNU Radio EN 50561 based application, MATLAB IEEE 1901-2010 based application) of software-defined power line communication (SD-PLC) systems. In \cite{b23}, Gary et al. presented in a detailed manner the implementation as well as a field test results of a software defined PLC (SD-PLC) modem utilizing digital signal processor (DSP). The developed SD-PLC modem supports multiple standard solutions such as G3-PLC, PRIME and IEEE P1901.2 and could be configured to utilize several OFDM technologies for the purpose of combating PLC channel impairment unique to a certain region’s grid condition. The field test results obtained from multiple test sites demonstrated and established the benefits of the flexible SD-PLC modem configuration in the improvement of throughput and coverage.

\section{Modeling Methodology}
\label{blockdiagonal}
HMMs are the class of Markov model utilized in the description of burst error in digital communication channels, where the burst error source state is presupposed to be in one of many states, the error probability is state dependent, with the error sequence observable. The name HMM is as a result of the unobservable underlying state sequence. HMMs are depicted by a state transition probability matrix $A$, prior state probability vector $p$ and the output symbol probability matrix $B$. Utilizing state splitting technique \cite{b13}, the errors could be made deterministic and not a probabilistic function of the state, thus resulting in more states, but with the $B$ matrix containing binary entries of zeros and ones only. Moreover, assuming a stationary model, subsequently, $p$ is defined by $pA = p$. Therefore, a stationary Markov model for burst errors can be depicted by the $A$ of suitable dimension. Henceforth, such reduced models are considered. 

Several techniques have been proposed for HMM parameter estimation, with the most widely used being the Baum-Welch iterative re-estimation algorithm \cite{b24}. Gradient-search principle, interval distribution curve fitting techniques have also been found to be useful, with the later computationally fast nonetheless, it does not depict  all the parameters of the HMM. The fundamental issue in the use of HMM is the estimation of $A$ from a simulated or empirically given error sequence. Simple estimation algorithms are obtainable should certain restrictions be imposed on the model structure. Although iterative in nature, the Baum-Welch algorithm is however computationally intensive given error sequence with long stretch of identical symbols (1's and 0's). Thus, in this paper, a faster procedure for the estimation of Markov models for burst error channels from a set of empirically obtained error sequences is adopted based on the technique in \cite{b14}. It is first proven, that for a Markov model with random state transition matrix $A$, there exist a unique equivalent ``\emph{Block Diagonal Markov Model}'' with state transition matrix $\Lambda$ for a model specified and represented by its state transition matrix. The two HMM $A$ and $\Lambda$ are determined to be equivalent should the probability of each empirical error sequence $E$ be the same for $A$ and $\Lambda$, i.e. $P(E|A) = P(E|\Lambda) \forall E$. A modified Baum-Welch algorithm (BWA) for estimation of $\Lambda$ is developed. Due to the uniqueness of the “Block Diagonal” nature of $\Lambda$, the computation in the modified BWA through the long strings of identical error sequence symbols is reduced to only the computation of the powers of the diagonal matrix and not the power of the random matrix, thus leading to computational savings. The realization of the equivalent model is constrained to some mild conditions on $A$, which holds consistent for most practical error sources.   

\begin{figure}[htbp]
\centerline{\includegraphics[width=0.40\textwidth]{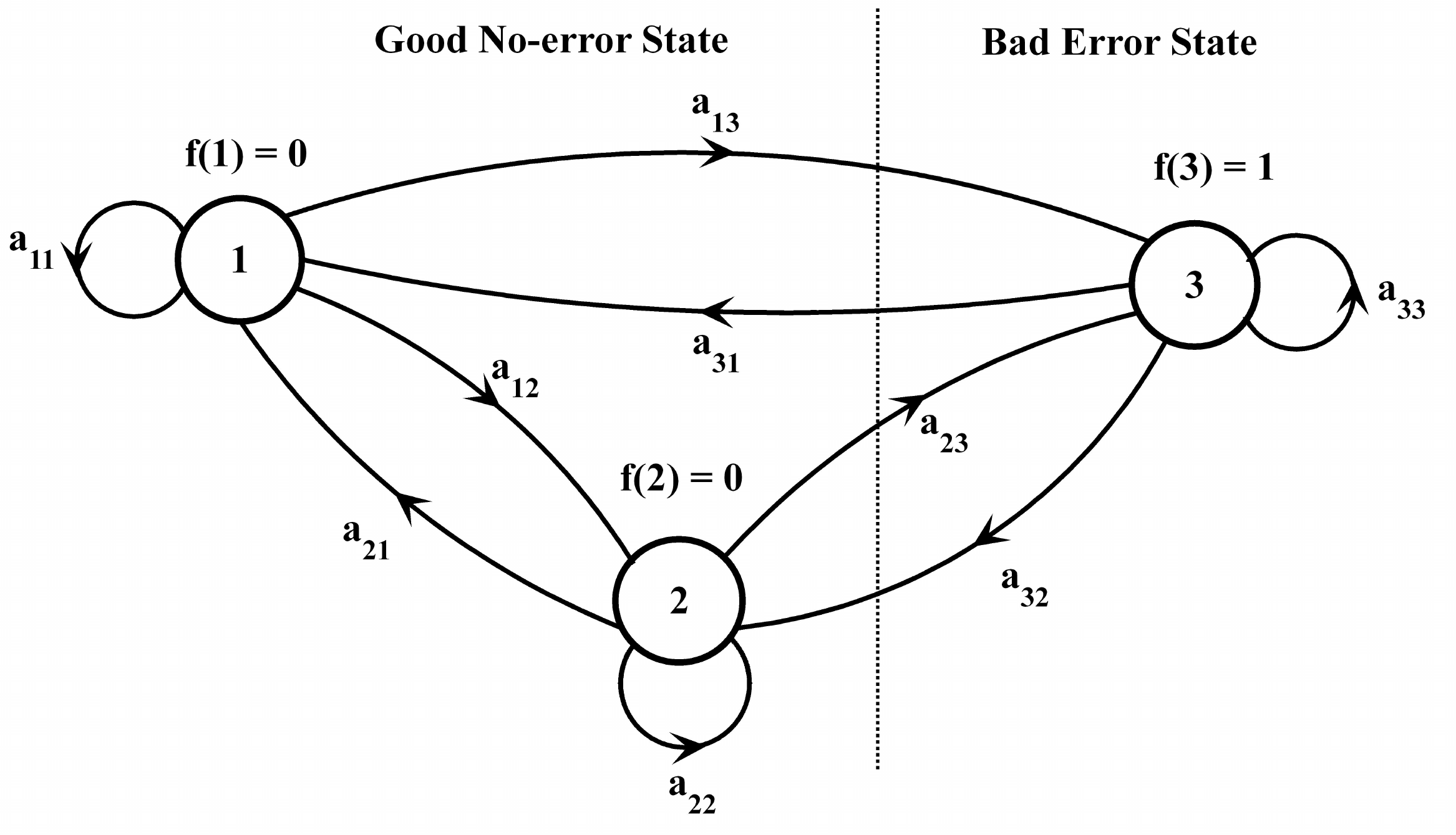}}
\caption{The initial Markov model $A$ partitioned ($N$ = 3).}
\label{GenSHMM}
\end{figure}

Fig.~\ref{GenSHMM} shows the state splitting of the Markov model, showing two states partition grouping which corresponds to two symbol $\mathcal{M} = \{0, 1\}$. ``0'' typifies an error free state (depicting no transmission error) and ``1'' typifies an error state (depicting a transmission error). Therefore, $n(0) = 2, n(1) = 1$ and $N = n(0) + n(1)$. Refer to Appendix for the construction of the Block Diagonal Markov model notations, construction and equivalence of $A$ and $\Lambda$. The equivalent Markov model for Fig.~\ref{GenSHMM} is shown in Fig.~\ref{EqSHMM}.

\section{The Modified Baum-Welch Algorithm}
\label{modbaum}
A faster and computationally very efficient modified Baum-Welch algorithm for estimating $\Lambda$ is presented in this section. For a distinct case of $\Lambda = A$, this algorithm is same as that proposed in \cite{b13}, but can generally be utilized in modeling source models that fulfil the constraints highlighted in III.B but are else random, while on the contrary the procedure proposed in \cite{b13} is computationally complex should transition between similar states be permitted.  
Given an empirically obtained error sequence $E$ of length $T$, the modified BWA exploit the benefit of the distinct structure of $\Lambda$ in computing $P(E|\Lambda)$. Due to $\Lambda_{\epsilon\epsilon}$ being diagonal matrices, computing $P(E|\Lambda)$ is considerably faster than computing $P(E|A)$ particularly when $E$ contains a long strings of identical error symbols. Let $m(\mu_{i})$ denote the number of consecutive occurrences of the symbol $\mu_{i}$ in $E$. Therefore, $E$ can be written as follows: $E = \mu_{1}^{m(\mu_{1})} \mu_{2}^{m(\mu_{2})} \mu_{3}^{m(\mu_{3})} \mu_{4}^{m(\mu_{4})} \cdots  \mu_{C}^{m(\mu_{c})}$.

For instance, given a sample error sequence as follows. $E = 00011000000111110110010000011001000$. It can be re-written in shorthand form as $E = 0^3 1^2 0^6 1^5 0^1 1^2 0^2 1^1 0^5 1^2 0^2 1^1 0^3$. Refer to Appendix for Baum-Welch algorithm derivations and how the modified Baum-Welch is used to estimate $\Lambda$. Refer to Appendix~\ref{AppendixA}, \ref{AppendixB}, \ref{AppendixC} for the block diagonal Markov model notation, construction, equivalence of \emph{A} and $\Lambda$ respectively. Appendix~\ref{AppendixD} presents the modified Baum-Welch algorithm derivation.

\section{System Model and Experiment}
\begin{figure}[htbp]
\centerline{\includegraphics[width=0.50\textwidth]{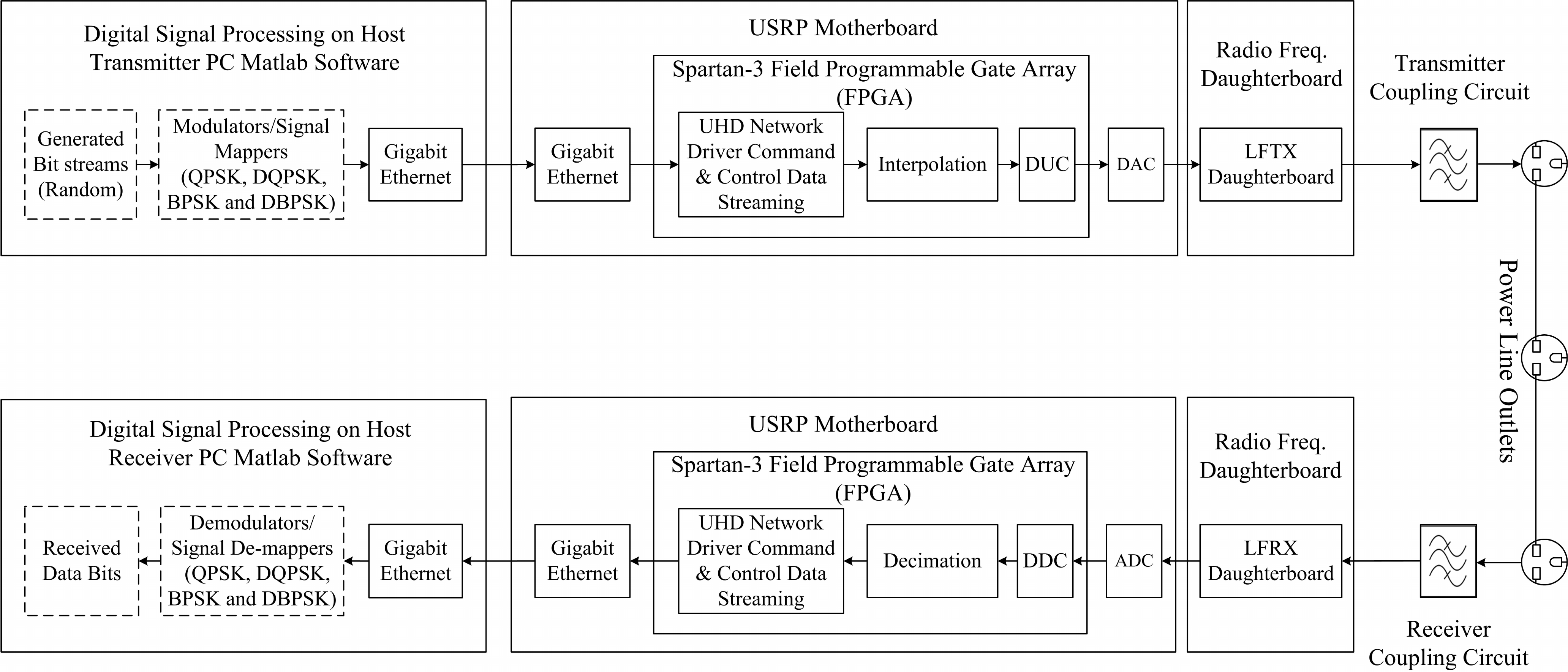}}
\caption{The software-defined single-carrier NB-PLC transceiver system model.}
\label{SDTransc}
\end{figure}

\begin{figure}[htbp]
\centerline{\includegraphics[width=0.50\textwidth]{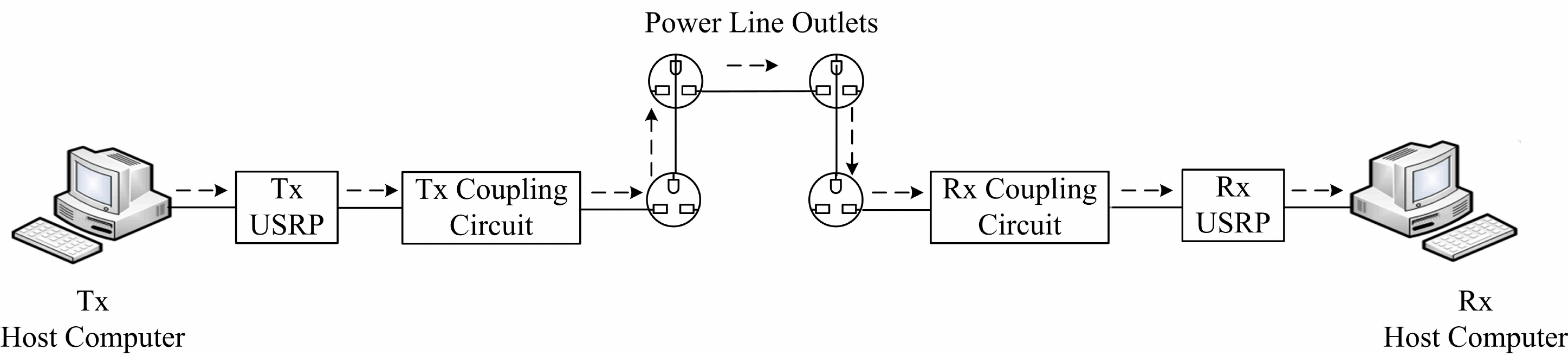}}
\caption{Architecture of the NB-PLC transceiver testbed.}
\label{SDArch}
\end{figure}

{}
A software-defined single-carrier NB-PLC transceiver system is developed using a Quadrature Phase Shift Keying (QPSK) digital modulation for real time signal transmission. Fig.~\ref{SDTransc} shows the system model for the NB-PLC transceiver. In Fig.~\ref{SDTransc} communication blocks depicted with dotted lines are software-based. The input signal of the QPSK system is modulated and mapped onto the constellation points of the QPSK utilizing the corresponding QPSK modulator. Fig.~\ref{SDArch} shows the architecture of the PLC transceiver testbed.

On the transmitter (Tx) side, the transmitting universal software radio peripheral (USRP) obtains the pre-processed modulated signal from the host Tx computer via the gigabit Ethernet interface and then stores it in a buffer for baseband processing. The samples then goes through interpolation, up-sampling actions before being modulated to the intermediate frequency (IF). The samples are then interpolated, up-sampled and then modulated to the IF from the buffer by sine and cosine wave for the complex samples, and by cosine waves for the real samples. Fundamentally, the digital up-converter (DUC) converts the digital complex baseband signal to real digital passband signal \cite{b25}. The DUC performs pulse shaping function on the incoming signal and also modulates it to a suitable intermediate carrier frequency appropriate for driving a final analog up-conversion \cite{b25}. The digital-to-analog converter (DAC) then transforms the digital signal into analog signal suited for transmission on the PLC channel prior to sending it to the LFTX daughterboard. The USRP LFTX daughterboard performs modulation of the Tx streams from the IF to the NB-PLC operational transmit frequency. The resulting continuous analog signal is then superimposed onto the voltage waveform of the NB-PLC channel by the Tx coupling interface.

On the receiver (Rx) side, the transmitted analog signal is decoupled from the PLC channel by the Rx coupling interface and passed to the USRP through the LFRX daughterboard. The LFRX daughterboard filters and modulate the received analog signals streams from the operational Rx frequency to the IF by initially sampling and then multiplying with a discrete time sine and cosine for complex samples and by cosine wave for real samples \cite{b26}. Afterward, the analog-to-digital converter (ADC) transforms the signal from analog form to digital bit form for subsequent baseband processing. The resulting digital bit stream is additionally filtered by the digital down-converter (DDC) and decimated so as to eliminate the dual frequency parts and decrease the sample rate. The resulting baseband signal samples will then be stored in the buffer and forwarded to the host Rx PC via the gigabit Ethernet interface for further baseband signal post-processing \cite{b25}. Lastly, the baseband signals is demodulated by the QPSK demodulator  consistent with the modulator utilized at the transmitter side recovering the original sent signal correctly in the absence of noise impairments or incorrectly in the presence of noise impairments. Error sequences are then obtained in the laboratory measurement site based on comparison between the sent and received signal with ``0'' denoting no transmission error, while ``1'' denotes an occurrence of transmission error. 

\section{Model Results and Analysis}
\begin{figure}[htbp]
\centerline{\includegraphics[width=0.40\textwidth]{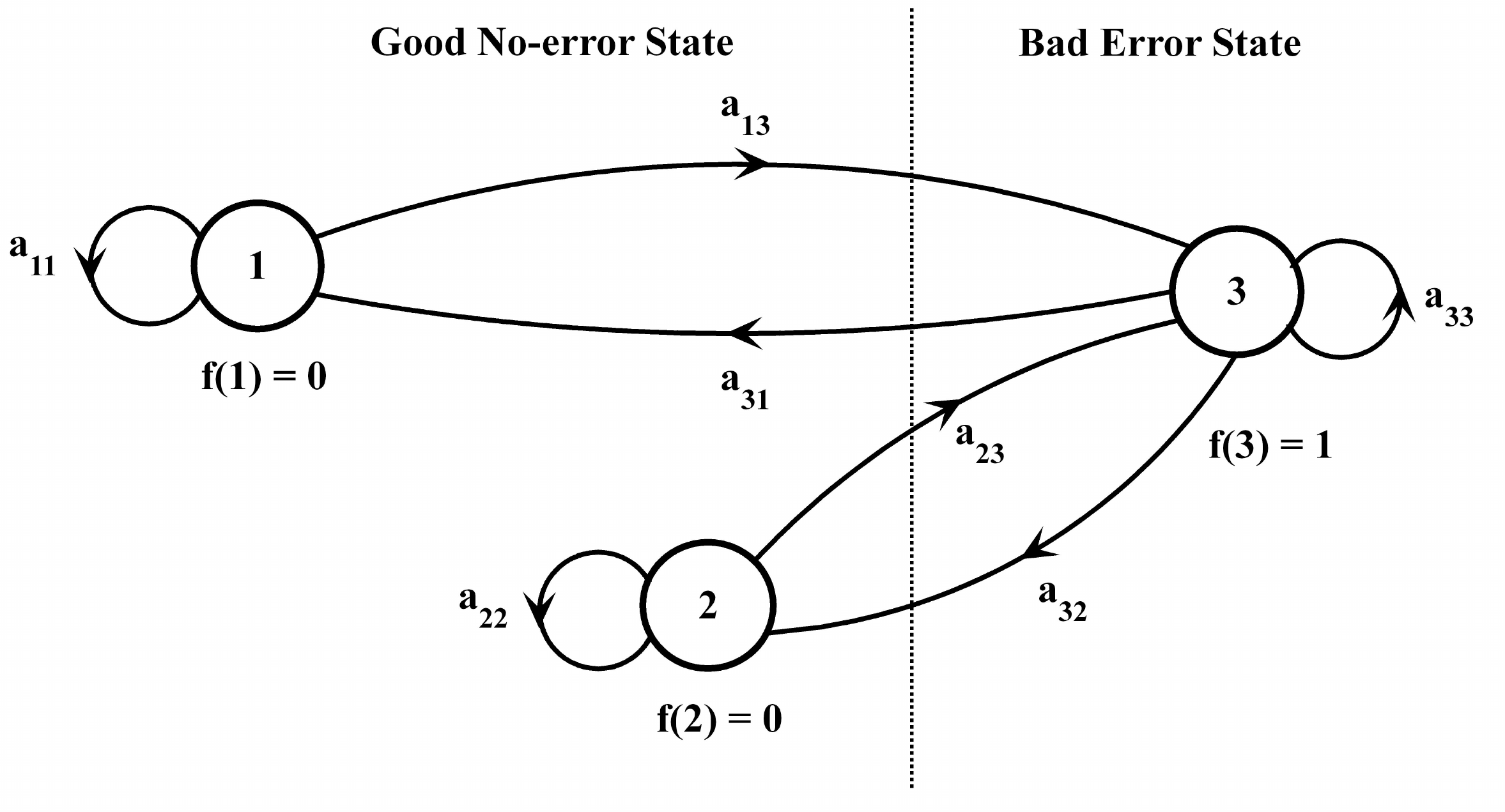}}
\caption{The structure of the estimated equivalent Markov model $\Lambda$ (N=3).}
\label{EqSHMM}
\end{figure}

Fig.~\ref{EqSHMM} shows the structure of the estimated models $\Lambda$ for the Markov model $A$ in Fig.~\ref{GenSHMM}. 

Since the good states are synonymous to error-free transmission, while errors are always obtained in the bad state, the output symbol error probability matrix $B$ is of binary format as follows.

\begin{equation*}
{\bf B} = \begin{bmatrix}
1 & 1 & 0   \\
0 & 0 & 1\end{bmatrix}.
\end{equation*} 

The assumed model is a three state Markov model and is semi-hidden, since if there is an occurrence of an error, we are quite sure it was generated by state three (the bad state). But, for a no error occurrence, we are unable to identify the state that generated it from the two good states. In a laboratory environment, error sequences of length 20,000 are obtained over the period of two days based on experimental measurement and taking into consideration the time of day \emph{morning}, \emph{afternoon} and \emph{evening}. We then train the modified Baum-Welch algorithm with this error sequences in order to obtained the estimated equivalent state transition matrix $\Lambda$ for each error sequence obtained.  

Table~\ref{InitialEst} shows the initial estimates $\hat \Lambda_{0}$, while Table~\ref{BlockEst} shows the re-estimated state transition probabilities values that depicts the measured error sequence. In other words, they are the most probable state transition probabilities values that produced the measured error sequences. It can be deduced from Table~\ref{BlockEst}, that the re-estimated state transition probabilities values are non-uniform, which is attributed to the fact that no two error sequence can be the same due to different channel status at the time of measurement. The usefulness of these model parameters is that, it can be used to facilitate the design and performance evaluation of forward error correcting codes useful in mitigating noise and resulting errors on the PLC channel. The model parameters can also be used to design adaptive modulation techniques, matching a modulation scheme based on the channel conditions. 

\begin{table}[!htb]
        \renewcommand{\arraystretch}{0.92}
        \caption{Initial estimates $\hat\Lambda_{0}$ for the measured Error sequence obtained over two days (Morning, Afternoon and Evening)}\label{InitialEst}
        \begin{center}
        \begin{tabular}{c|c|c|c|c|c|c}
                \hline\hline
                                &\multicolumn{3}{c|}{Day I} & \multicolumn{3}{c}{Day II}   \\
               \hline
                                &  Morning  &  Noon     &  Evening   &  Morning  & Noon      &  Evening  \\
              \hline
               $\Lambda_{11}$   &  0.9000   &  0.9200   &  0.9215    &  0.9266   &  0.9539   &  0.9794  \\
               $\Lambda_{13}$   &  0.1000   &  0.0800   &  0.0785    &  0.0734   &  0.0461   &  0.0206  \\
               $\Lambda_{22}$   &  0.9200   &  0.9220   &  0.9346    &  0.9224   &  0.9229   &  0.9628  \\
               $\Lambda_{23}$   &  0.0800   &  0.0780   &  0.0654    &  0.0776   &  0.0771   &  0.0372  \\
               $\Lambda_{31}$   &  0.1200   &  0.0085   &  0.7285    &  0.5000   &  0.3856   &  0.4021  \\
               $\Lambda_{32}$   &  0.7000   &  0.9364   &  0.2060    &  0.4200   &  0.5711   &  0.5008  \\
               $\Lambda_{33}$   &  0.1800   &  0.0551   &  0.0655    &  0.0800   &  0.0433   &  0.0971  \\
              \hline\hline
        \end{tabular}
        \end{center}
\end{table}

\begin{table}[!htb]
        \renewcommand{\arraystretch}{0.92}
        \caption{Converged model estimates $\hat \Lambda$ for the measured Error sequence obtained over two days (Morning, Afternoon and Evening)}\label{BlockEst}
        \begin{center}
        \begin{tabular}{c|c|c|c|c|c|c}
                \hline\hline
                                &\multicolumn{3}{c|}{Day I} & \multicolumn{3}{c}{Day II}   \\
               \hline
                                &  Morning  &  Noon     &  Evening   &  Morning    & Noon     &  Evening  \\
              \hline
               $\Lambda_{11}$   &  0.9683   &  0.9810   &  0.9705    &  0.9564   &  0.9539   &  0.9594  \\
               $\Lambda_{13}$   &  0.0317   &  0.0190   &  0.0295    &  0.0436   &  0.0461   &  0.0406  \\
               $\Lambda_{22}$   &  0.9205   &  0.9229   &  0.9346    &  0.9726   &  0.9529   &  0.9428  \\
               $\Lambda_{23}$   &  0.0795   &  0.0771   &  0.0654    &  0.0274   &  0.0471   &  0.0572  \\
               $\Lambda_{31}$   &  0.1273   &  0.0585   &  0.2080    &  0.5098   &  0.3453   &  0.4501  \\
               $\Lambda_{32}$   &  0.8188   &  0.9064   &  0.7665    &  0.4624   &  0.5711   &  0.5021  \\
               $\Lambda_{33}$   &  0.0539   &  0.0351   &  0.0255    &  0.0278   &  0.0836   &  0.0478  \\
              \hline\hline
        \end{tabular}
        \end{center}
\end{table}

\begin{figure}[htbp]
\centerline{\includegraphics[width=0.53\textwidth]{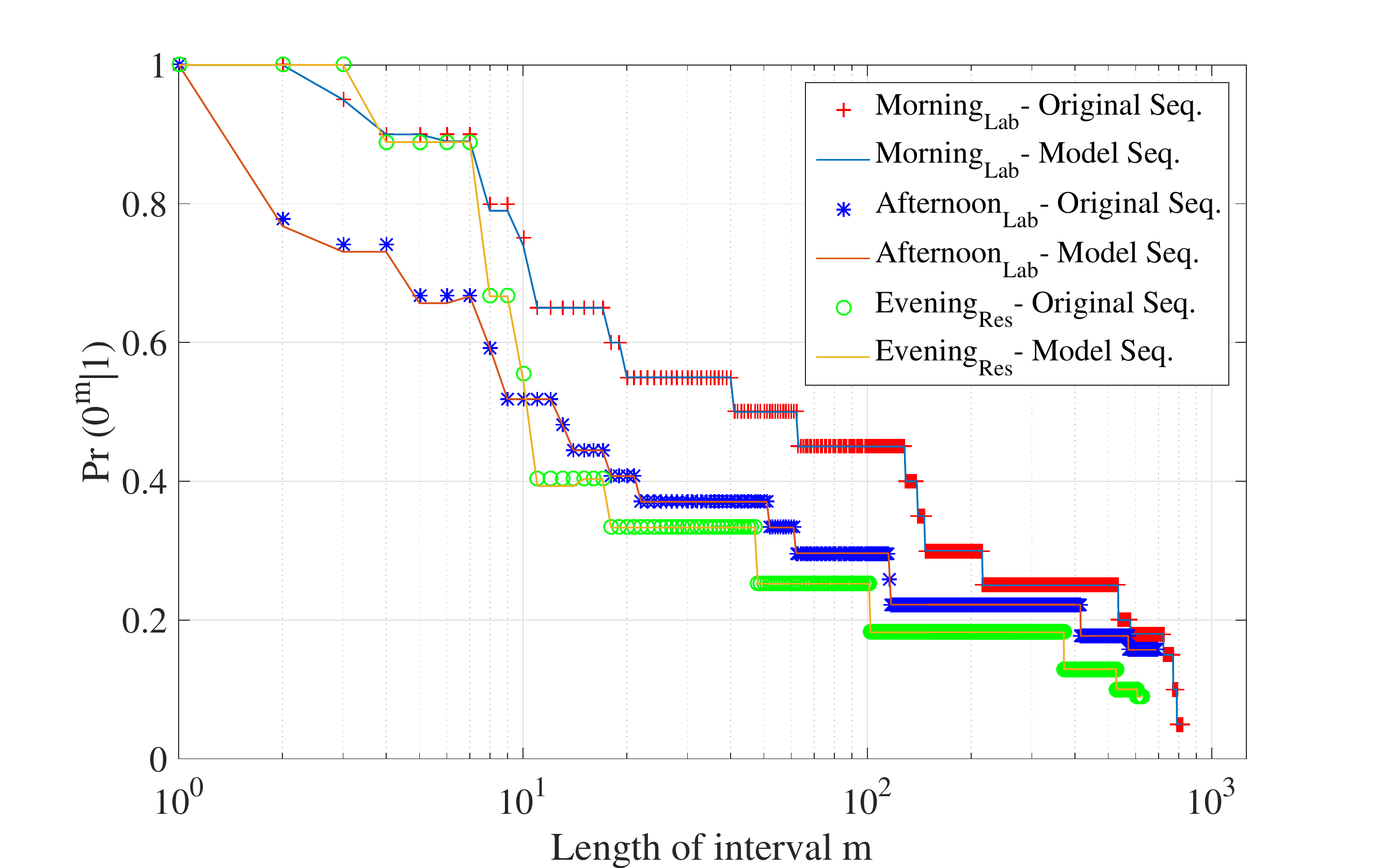}}
\caption{Error-free run distribution plot $\Pr (0^m|1)$ (Day I).}
\label{efrdI}
\end{figure}

\begin{figure}[htbp]
\centerline{\includegraphics[width=0.53\textwidth]{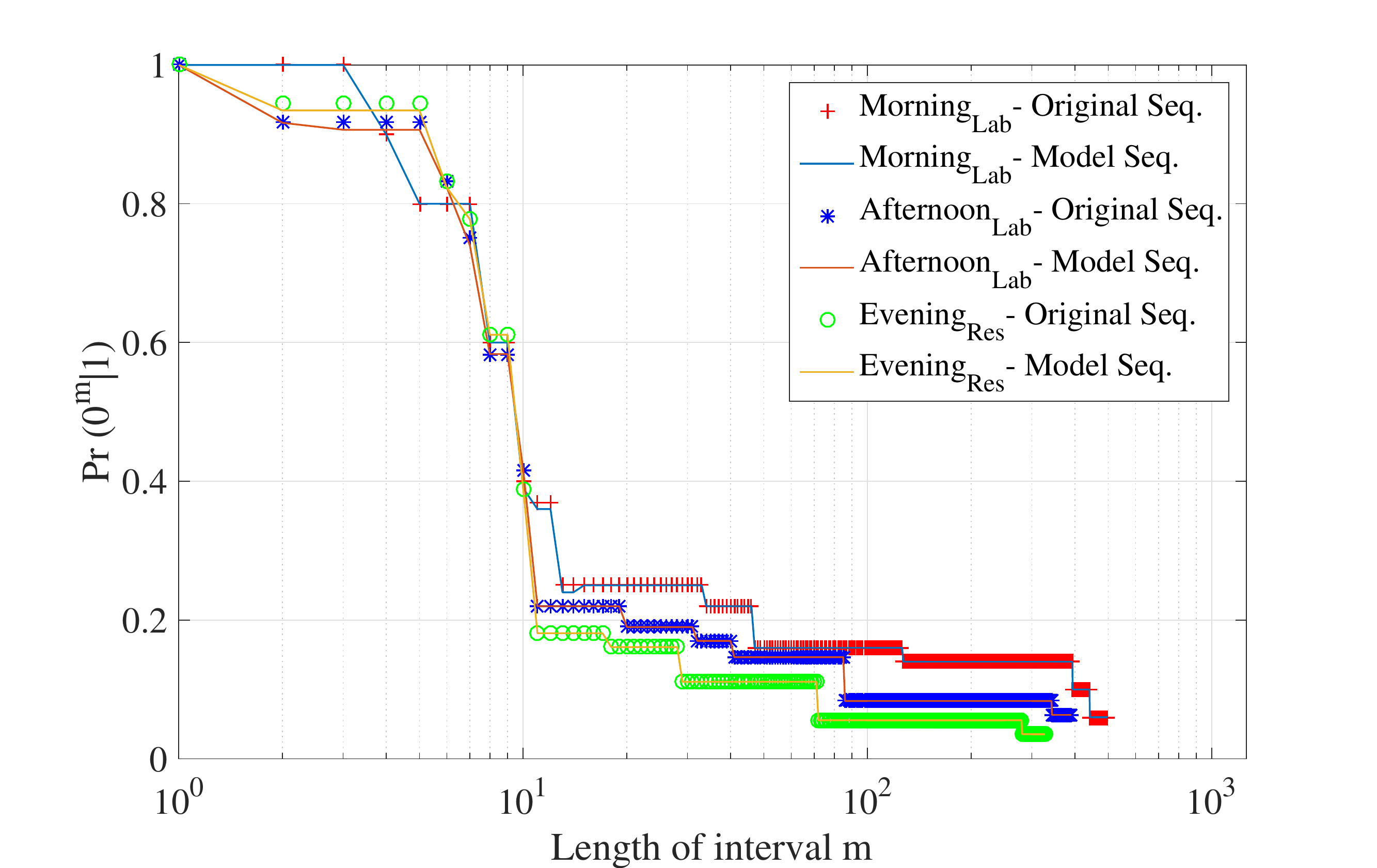}}
\caption{Error-free run distribution plot $\Pr (0^m|1)$ (Day II).}
\label{efrdII}
\end{figure}

In order to validate the precision of these models, a comparison between the error-free run distribution (EFRD) plot of the measured error sequence and that of the model generated error sequence is carried out. The error-free run distribution denoted by $\Pr (0^m|1)$ implies the probability of having $m$ consecutive error-free runs after a transition from an error state. A precise model should show a close agreement between the error-free run distribution plot of the measured error sequences and that of the model generated error sequences. Fig.~\ref{efrdI} and Fig.~\ref{efrdII} shows the EFRD plot for day I and day II respectively.  

It can be deduced from Fig.~\ref{efrdI} and Fig.~\ref{efrdII}, that there is a close agreement between the error-free run distribution plot of the measured error sequences and the model generated error sequences.

Another way of validating the precision and fitness of the estimated model is to compare the error probabilities of the measured error sequence ($P_{e}$) with that of the model generated error sequence $\bar{P}_{e}$, with a close agreement validating the estimated model. Table~\ref{errorprob} show this comparison, and it can be deduced that there is a close agreement between the error probabilities of the measured error sequences with the model generated error sequences. 

\begin{table}[!htb]
        \renewcommand{\arraystretch}{1.0}
        \caption{Error probabilities for measured original error sequence ($P_{e}$) and model regenerated error sequence ($\bar{P}_{e}$)- First-Order SHFMM}\label{errorprob}
        \begin{center}
        \begin{tabular}{c|c|c|c|c|c|c}
                \hline\hline
                               &\multicolumn{3}{c|}{Day I} & \multicolumn{3}{c}{Day 2} \\
                 \hline
                               &  Morning   &  Noon  &  Evening &  Morning   & Noon   &  Evening\\
                \hline
                 $P_{e}$       &  0.0512  &  0.0330  &  0.0762  &  0.0487  & 0.0389  &  0.0678\\
                 $\bar{P}_{e}$ &  0.0501  &  0.0320  &  0.0751  &  0.0476  & 0.0378  &  0.0667\\
                \hline\hline
        \end{tabular}
        \end{center}
\end{table}

\section{Conclusion}
In this paper, a block diagonal Markov model is utilized in modeling a software-defined based PLC system in an indoor environment. A computationally efficient modified \emph{Baum-Welch algorithm} is used to obtained the estimated model $\hat \Lambda$ for the measured error sequences. Results shows a close agreement between the measured error sequence and estimated model generated error sequence validating the fitness and precision of the model. The resulting estimated models can be used to facilitate the design and performance evaluation of forward error correcting codes useful in mitigating noise and resulting errors on the PLC channel. The model parameters can also be used to design adaptive modulation techniques, which matches a robust modulation scheme based on the PLC channel conditions.

\vspace{12pt}
%\color{red}\textbf{Note that the Appendix will be uploaded on arXiv as a separate supplementary file for this paper or on any IEEE repository suggested by the organizers.} 

\newpage
\color{black}
\appendix
\label{Appendix}

\subsection{Block Diagonal Markov Model Notation}
\label{AppendixA}
Let $\mathcal{M} = \{\epsilon_{1}, \epsilon_{2}, \epsilon_{3}, \cdots ,\epsilon_{d}\}$ represent a set of observed symbols, while $\epsilon, \mu, \mu_{1}, \mu_{2}, \mu_{3}, \cdots$  are variables drawing values from $\mathcal{M}$. $E$ denotes an empirically obtained error sequence, while $E$ with subscripts denotes particular instances of $E$.  Let $N$ be the number of model states and $n(\epsilon)$ represents the number of states from which symbol $\epsilon$ could be observed. Thus, $N = \sum_{\epsilon}n(\epsilon)$. Any state transition matrix $A$ has a $N \times N$ matrix structure that can be expressed in a block matrix form using sub-matrices of size $n(\epsilon) \times n(\mu)$. 

\begin{equation}
{\bf A} = \begin{bmatrix}
A_{{\epsilon}_1}{_{\epsilon_1}}  & A_{{\epsilon}_1}{_{\epsilon_2}} & \cdots & A_{{\epsilon}_1}{_{\epsilon_d}}\\
A_{{\epsilon}_2}{_{\epsilon_1}}  & A_{{\epsilon}_2}{_{\epsilon_2}} & \cdots & A_{{\epsilon}_2}{_{\epsilon_d}} \\
 & \vdots & \vdots & \\
A_{{\epsilon}_d}{_{\epsilon_1}}  & A_{{\epsilon}_d}{_{\epsilon_2}}  & \cdots & A_{{\epsilon}_d}{_{\epsilon_d}}
% \begin{equation}
\end{bmatrix}  
\label{Amatrix} %\eqref{Amatrix}
\end{equation} 
              
% {\bf A} = \begin{bmatrix}
% A_{\epsilon_{1}}_{\epsilon_{1}} & A_{\epsilon_{1}}_{\epsilon_{2}} &  A_{\epsilon_{1}}_{\epsilon_{3}}\\
% A_{\epsilon_{2}}_{\epsilon_{1}} & A_{\epsilon_{2}}_{\epsilon_{2}} & A_{\epsilon_{2}}_{\epsilon_{3}}\\
% A_{\epsilon_{3}}_{\epsilon_{1}} & A_{\epsilon_{3}}_{\epsilon_{1}} & A_{\epsilon_{3}}_{\epsilon_{3}}
% \end{bmatrix}  
% \label{Amatrix3} %\eqref{Amatrix3}
% \end{equation} 

The Markov model states $N$ are thus partitioned into $d$ groups. An assumption is made that the empirically obtained error sequence is a function of the Markov chain produced by the $A$ matrix. Therefore, identifying the group at any time from the empirically obtained error sequence is always possible, but not definitely the actual state in the group. The state splitting of the Markov model is depicted in Fig. 1 showing two states partition grouping which corresponds to two symbol $\mathcal{M} = \{0, 1\}$. ``0'' typifies an error free state (depicting no transmission error) and ``1'' typifies an error state (depicting a transmission error). Therefore, $n(0) = 2, n(1) = 1$ and $N = n(0) + n(1)$. 

Each observed symbol $E_{t}$ specified by $E_{t} = f(S_{t})$, with $S_{t}$ being the state at discrete time $t$ and $f:\{s_{1}, s_{2}, s_{3}, \cdots ,s_{N}\} \rightarrow \mathcal{M}$ is generally non-one-to-one function. Suppose we identify the model states by their indices, i.e. $s_{i} = i, i = 1, 2, 3, \cdots , N$ , we can also identify the observed symbols by their indices, thus $\epsilon_{k} = k-1, k = 1, 2, 3, \cdots , d$. The transition matrix $\Lambda$ thus represent the equivalent model. Note that row vector $p$ and $\pi$ is used to represent the stationary probability vectors corresponding to $A$ and $\Lambda$ respectively. 

Let $\lambda_{{\epsilon}_k}{_{\epsilon_k}}(1), \lambda_{{\epsilon}_k}{_{\epsilon_k}}(2), \lambda_{{\epsilon}_k}{_{\epsilon_k}}(3), \dots ,\lambda_{{\epsilon}_k}{_{\epsilon_k}}(n(\epsilon_{k}))$ denote the Eigen-values of $A_{{\epsilon}_k}{_{\epsilon_k}}$ , while $\alpha$, $\beta, \gamma$ denotes column vectors of probabilities. A pseudo Markov matrix $M$ is also defined to be a matrix that fulfils the condition $\sum_{j}M_{ij} = 1$ implying that each row of the pseudo Markov matrix sums up to 1.   
Suppose the eigenvectors of $A_{\epsilon \epsilon}$ is denoted by $E_{\epsilon \epsilon}(i)$ for $i = 1, \cdots n(\epsilon)$ and $\epsilon = \epsilon_{1}, \cdots ,\epsilon_{d}$. Thus

\begin{equation}
A_{\epsilon\epsilon} = E_{\epsilon\epsilon} \Lambda_{\epsilon\epsilon} E_{\epsilon\epsilon}^{-1}.
\label{Alambda}
\end{equation}

with $E_{\epsilon \epsilon} = (E_{\epsilon \epsilon}(1) ~E_{\epsilon \epsilon}(2) ~E_{\epsilon \epsilon}(3) \cdots E_{\epsilon \epsilon}(n(\epsilon)))$ yielding the diagonal matrix $\Lambda_{\epsilon \epsilon}$ in a way that $\Lambda_{\epsilon \epsilon}(i,i) = \lambda_{\epsilon \epsilon}(i)$ is the ith eigenvalue of $A_{\epsilon \epsilon}$.  Allowing $V_{\epsilon \epsilon} = E_{\epsilon \epsilon}^{-1}$  and making $C_{\epsilon \epsilon}$ a diagonal matrix of normalizing constants resulting  in the following matrix.
\begin{equation}
{\bf C_{\epsilon \epsilon}} = \begin{bmatrix}
\sum V_{\epsilon \epsilon}(1,i) & 0 &  \cdots & 0\\
 & \ddots &  & \\
  &  & \ddots & \\
0 & 0 & \cdots & \sum V_{\epsilon \epsilon}(n(\epsilon),i)
\end{bmatrix}.  
\label{Cmatrix} %\eqref{Cmatrix}
\end{equation}   

Suppose that the diagonal matrix of normalizing constants $C_{\epsilon \epsilon}$ is non-singular, we then define $W_{\epsilon \epsilon} = V_{\epsilon \epsilon} C_{\epsilon \epsilon}^{-1}$.

\subsection{The Block Diagonal Markov Model Construction}
\label{AppendixB}
The state transition matrix can take the form stated in \eqref{Amatrix}with the block matrix $A_{\epsilon_{i} \epsilon_{j}}$ of a $n(\epsilon_{i} \times n(\epsilon_{j})$ dimension and comprising the transition probabilities from the $n(\epsilon_{i})$ states from which the observed error symbol $\epsilon_{i}$ emanated, to the $n(\epsilon_{j})$ states from which the observed error symbol $\epsilon_{j}$ emanated. Suppose that the block matrices $A_{\epsilon \epsilon}$ across the diagonal of $A$ meets the five conditions listed as follows, we will have a large enough collection of matrices satisfying these constraints which are of interest and such matrices will only be considered \cite{b14}.

\begin{enumerate}[i)]
\item $A_{\epsilon \epsilon}(j,j) \geq \sum_{k\neq j} A_{\epsilon \epsilon}(j,k)$, i.e $Re(\lambda_{\epsilon\epsilon})$ $>0$.
\item For all pairs of rows of $A_{\epsilon \epsilon}$ having indices $r$ and $s$ in a manner that $A_{\epsilon \epsilon}(s,s) > A_{\epsilon \epsilon}(r,r)$, suppose $\sum_{i=1}^{n(\epsilon)} A_{\epsilon\epsilon}(r,i) \leq A_{\epsilon\epsilon}(s,s) - \sum_{i\neq s}A_{\epsilon\epsilon}(s,i)$.
\item Assume $C_{\epsilon\epsilon}$ to be non-singular. 
\item All the elements of $W_{\epsilon\epsilon}^{-1} A_{\epsilon\mu} W_{\mu\mu}$ must be positive $\forall \epsilon \neq \mu$.
\item All matrices $A_{\epsilon\epsilon}$ must be a non-degenerate (or non-singular).
\end{enumerate}

The first conditional physically means that the transition probability self-transition to same state is higher than the probability of transitioning to any other state with similar symbol. The second condition makes the states adequately \emph{distinct} from one another. The third condition ensures that no non-essential states exist. Let’s create the block diagonal matrix $W$ as follows.

\begin{equation}        
{\bf W} = \begin{bmatrix}
W_{{\epsilon}_1}{_{\epsilon_1}} & 0 & \cdots & 0\\
0 &  W_{{\epsilon}_2}{_{\epsilon_2}} & \cdots & 0\\
 & \vdots & \vdots & \\
0 & 0 & \cdots & W_{{\epsilon}_d}{_{\epsilon_d}}
\end{bmatrix}  
\label{Wmatrix} %\eqref{Wmatrix}
\end{equation} 

and also define $\Lambda$ in terms of $W$ as follows.

\begin{equation}
A = W^{-1}\Lambda W
\label{A}
\end{equation} 

Thus it can be shown that $\Lambda$ also has the form.

\begin{equation}
\Lambda = \begin{bmatrix}
\Lambda_{{\epsilon}_1}{_{\epsilon_1}} & \Lambda_{{\epsilon}_1}{_{\epsilon_2}} & \cdots & \Lambda_{{\epsilon}_1}{_{\epsilon_d}}\\
\Lambda_{{\epsilon}_2}{_{\epsilon_1}} & \Lambda_{{\epsilon}_2}{_{\epsilon_2}} & \cdots & \Lambda_{{\epsilon}_2}{_{\epsilon_d}}\\
& \vdots & \vdots &\\
\Lambda_{{\epsilon}_d}{_{\epsilon_1}} & \Lambda_{{\epsilon}_d}{_{\epsilon_2}} & \cdots & \Lambda_{{\epsilon}_d}{_{\epsilon_d}}
\end{bmatrix}  
\label{Lmatrix3} %\eqref{Lmatrix3}
\end{equation} 

with $\Lambda_{{\epsilon}_i}{_{\epsilon_i}}$ being diagonal matrices for $i = 1, 2, 3, \cdots ,d$. Moreover, the four conditions on $A$ will guarantee that  is non-negative (positive) matrix and therefore $\Lambda$ is a stochastic matrix as shown as follows.
Suppose $\overrightarrow{1} = (1~ 1 ~ \cdots ~1)^{T}$ be the $N$ dimensional unity vector.
Therefore,

\begin{equation}
WA\overrightarrow{1} = W\overrightarrow{1} = \Lambda W\overrightarrow{1}.
\end{equation}

Because each row of $W$ sums up to unity, $W\overrightarrow{1} = \overrightarrow{1}$. Thus
\begin{equation}
\overrightarrow{1} = W\overrightarrow{1} = \Lambda W\overrightarrow{1} = \Lambda \overrightarrow{1}.
\end{equation}

$\pi$, the stationary probability distribution vector of $\Lambda$ is thus expressed by the expression $\pi = pW^{-1}$ since

\begin{equation}
pAW^{-1} = pW^{-1} = pW^{-1} \Lambda.
\end{equation}

The probabilities value in $\pi$ sums up to 1 since

\begin{equation}
\sum_{i}\pi_{i} = \sum_{j} \sum_{i}\pi_{i}W_{ij} = \sum_{i}p_{i} = 1.
\end{equation}

Because the underlying Markov chain is taken to be regular, $\pi$ is the stationary probability distribution of $\Lambda$. \eqref{Alambda}, \eqref{Cmatrix}, \eqref{Wmatrix}, \eqref{A}, \eqref{Lmatrix3} show the procedure for construction of $\Lambda$, but this is utilized only to establish that $\Lambda$ exist, is equivalent and unique. In practice, it should be noted that $\Lambda$ is not obtained from \eqref{A}since $A$ is unknown. Thus, rather than estimate $A$ from a given empirical error sequence, one estimates $\Lambda$ utilizing the computationally efficient algorithm derived in Section IV and utilize $\Lambda$ as the model for the NB-PLC channel. Statistically, the use of $\Lambda$ as a model is equivalent to the use of $A$, with $A$ very monotonous to obtain.  

\subsection{Equivalence of \emph{A} and $\Lambda$}
\label{AppendixC}
The HMM determined by $A$ and $\Lambda$ are believed to be equivalent if for each and every empirically obtained error sequence $E$, the likelihood of $E$ is equivalent under $A$ and $\Lambda$, i.e. $P(E|A) = P(E|\Lambda)$. The equality of $A$ and $\Lambda = WPW^{-1}$ shown in mathematical form as follows [14]:
Let $E=\{\mu_{1}, \mu_{2}, \cdots ,\mu_{n}\}$ be any empirically obtained error sequence. 

\begin{eqnarray*}
P(E|A) = p_{\mu_{1}}A_{\mu_{1}\mu_{2}}A_{\mu_{2}\mu_{3}} \cdots A_{\mu_{n-1}\mu_{n}}\overrightarrow{1}\\
       = p_{\mu_{1}}W_{\mu_{1}\mu_{1}}^{-1}\Lambda_{\mu_{1}\mu_{2}}\Lambda_{\mu_{2}\mu_{3}} \cdots \Lambda_{\mu_{n-1}\mu_{n}}W_{\mu_{n}\mu_{n}}\overrightarrow{1}\\
       = \pi_{\mu_{1}}\Lambda_{\mu_{1}\mu_{2}}\Lambda_{\mu_{2}\mu_{3}} \cdots \Lambda_{\mu_{n-1}\mu_{n}}\overrightarrow{1}\\
       = P(E|\Lambda)
\end{eqnarray*}

considering $W_{\mu_{n}\mu_{n}}\overrightarrow{1} = \overrightarrow{1}$. Therefore $A\equiv \Lambda$. Because $W$ is distinctively defined in terms of the eigenvectors of $A_{\epsilon\epsilon}$, (5) is a distinctive transformation that takes $A$ to $\Lambda$. Nevertheless, there could exist more than one source model $A$, for instance $A_{1}$ and $A_{2}$  in such a way that $P(E|A_{1}) = P(E|A_{2})$. Thus, a bijection between $A$ and $\Lambda$ does not sufficiently guarantees a distinct $\Lambda$ in terms of the statistics of the empirically observed error sequences. Nonetheless, a proof showing that $\Lambda$ is distinctly specified by the statistics of the empirically observed error sequences is shown in \cite{b14}.

\subsection{The Modified Baum-Welch Algorithm Derivation}
\label{AppendixD}
We define the $N \times 1$ vectors $\alpha_{c}$ and  $\beta_{c}$ as follows 

\begin{equation*}
\alpha_{c} = \begin{pmatrix}
P(\mu_{1}^{m(\mu_{1})} \cdots \mu_{c}^{m(\mu_{c})}, q_{t} = 1|\Lambda)  \\
P(\mu_{1}^{m(\mu_{1})} \cdots \mu_{c}^{m(\mu_{c})}, q_{t} = 2|\Lambda)\\
        \vdots\\
P(\mu_{1}^{m(\mu_{1})} \cdots \mu_{c}^{m(\mu_{c})}, q_{t} = N|\Lambda)\end{pmatrix}.
\end{equation*}

\begin{equation*}
\alpha_{c} = \begin{pmatrix}
P(\mu_{c+1}^{m(\mu_{c+1})} \cdots \mu_{C}^{m(\mu_{C})}, q_{t} = 1|\Lambda)  \\
P(\mu_{c+1}^{m(\mu_{c+1})} \cdots \mu_{C}^{m(\mu_{C})}, q_{t} = 2|\Lambda)\\
        \vdots\\
P(\mu_{c+1}^{m(\mu_{c+1})} \cdots \mu_{C}^{m(\mu_{C})}, q_{t} = N|\Lambda)\end{pmatrix}.
\end{equation*}

with $t = m(\mu_{1}) + \cdots + m(\mu_{c})$ and $q_{t}$ denoting the state at time $t$. Lastly the restriction of a vector to a subspace is defined. For instance $\alpha_{c|\mu}$ refers to the restriction of the vector $\alpha_{c}$ to the Euclidean subspace (of dimension $n(\mu)$) matching the symbol $\mu$, i.e. it denotes the sub-vector of elements matching the states of the symbol $\mu$. The Forward and Backward recursion equations utilized in computing the vectors $\alpha_{c}, \beta_{c}$ as well as the probability $P(E|\Lambda)$ is developed as follows.\\

\textbf{The Forward Iterative Computation}

\begin{enumerate}
\item Initialization:
\begin{equation*}
\alpha_{1|{\mu_{1}}} = \pi_{\mu_{1}} \Lambda_{{\mu_{c}}\mu_{c}}^{m(\mu_{c})-1}
\end{equation*}
\begin{equation*}
\alpha_{1|\epsilon\neq\mu_{1} = \overrightarrow{0}}
\end{equation*}

\item Induction: 
\begin{equation*}
\alpha_{c|{\mu_{c}}} = \alpha_{{c-1}|\mu_{c-1}} \Lambda_{{\mu_{c-1}}\mu_{c}} \Lambda_{{\mu_{c}}\mu_{c}}^{m(\mu_{c})-1}
\end{equation*}

\begin{equation*}
\alpha_{c|\epsilon\neq\mu_{c}} = \overrightarrow{0}, c =2, \cdots, C
\end{equation*}

\item Termination:
\begin{equation*}
P(E|\Lambda) = \sum_{i=1}^{N} \alpha_{C}(i) = \sum_{i_{\mu_{C}=1}}^{n{(\mu_{C})}}\alpha_{C|\mu_{C}}(i_{\mu_{C}})
\end{equation*}
\end{enumerate}

\textbf{The Backward Iterative Computation}
\begin{enumerate}
\item Initialization:

\begin{equation*}
\beta_{C} = \overrightarrow{1}
\end{equation*}

\item Induction:

\begin{eqnarray*}
\beta_{c|\epsilon} = \beta_{c+1|c+1}\Lambda_{\epsilon\mu_{c+1}}\Lambda_{\mu_{c+1}\mu_{c+1}}^{m(\mu_{c+1})-1} ~~\forall\epsilon = (\epsilon_{1}, \cdots, \epsilon_{d}) \\ \forall c = 1, \cdots, C-1.
\end{eqnarray*}
\end{enumerate}

To re-estimate the equivalent model $\Lambda$ utilizing Baum-Welch algorithm, $\gamma_{t}$  is defined as an $N \times 1$ vector and likewise, $\Gamma_{t}$ defined as an $N \times N$ matrix in a way that

\begin{eqnarray*}
\gamma_{t}(i) \triangleq P(q_{t} = i|E, \Lambda) ~~~\forall i \in (1,N)\\
\Gamma_{t}(i,j) \triangleq P(q_{t} = i, q_{t+1}=j|E, \Lambda)
\end{eqnarray*}

There is need to only define $\gamma$ and $\Gamma$ at $t=m(\mu_{1}) + \cdots + m(\mu_{c}), \forall c = 1, \cdots, C$ to be able to compute the matrix $\Lambda$. It can be substantiated that $\gamma_{c}$ and  $\Gamma_{c}$ can be expressed in terms of both the forward and backward variables $\alpha_{c}$ and $\beta_{c}$ as follows 

\begin{eqnarray*}
\gamma_{c}(i) = \frac{\alpha_{c}(i)\beta_{c}(i)}{P(E|\Lambda)}\\
\Gamma_{\mu_{c}\mu_{c+1}, c} = \frac{\alpha_{c|\mu_{c}}\Lambda_{\mu_{c}\mu_{c+1}}\Lambda_{\mu_{c+1}\mu_{c+1}}^{m(\mu_{c+1})-1}}{P(E|\Lambda)}\\
\Gamma_{\epsilon\mu} = [0] ~unless ~\epsilon=\mu_{c} ~and~ \mu = \mu_{c+1}.
\end{eqnarray*}
Where $[0]$ denotes a matrix made up of all zeros.\\ 

\textbf{The Re-estimation Formulas}
The BWA re-estimation procedures are stated as follows:

\begin{equation*}
\sum_{t=1}^{C-1}\gamma_{t}(i) = anticipated ~no.~ of~ transitions ~from~ state~ i
\end{equation*}

\begin{equation*}
\sum_{t=1}^{C}\gamma_{t}(i) = anticipated ~no.~ of~ times ~state~ i~is~visited
\end{equation*}

\begin{equation*}
\sum_{t=1}^{C-1}\Gamma_{t}(i,j) = anticipated ~no.~ of~ transitions ~from~ i~to~j
\end{equation*}

Therefore the modified Baum-Welch re-estimation formula are expressed as follows

\begin{equation*}
\tilde{\pi} = \gamma_{1}
\end{equation*}

\begin{equation*}
\tilde{\Lambda}_{\epsilon\epsilon}(i,j) = \frac{[\sum_{c=1}^{C}\Gamma_{\epsilon\epsilon,c}(i,j) + \sum_{c=1}^{C}\gamma_{c|\epsilon}(i)(m_{c}(\epsilon)-1)]\delta_{ij}}{\sum_{c=1}^{C-1}\gamma_{c}(i) + \gamma_{C}(i)(m_{c}(\epsilon)-1)}
\end{equation*}

note that, $f(i) = f(j) = \epsilon$

\begin{equation*}
\tilde{\Lambda}_{\epsilon\mu}(i,j) = \frac{\sum_{c=1}^{C-1}\Gamma_{\epsilon\mu,c}(i,j)}{\sum_{c=1}^{C-1}\gamma_{c}(i) + \gamma_{C}(i)(m_{c}(\epsilon)-1)}
\end{equation*}

Note that $\delta_{ij}$ denotes the krõnecker delta function. Scaling of the forward and backward variables computation is important at each transition time step. The scaling technique described in \cite{b14} can be utilized for the modified BWA to prevent numerical underflow. 
\end{document}